\documentclass[aps,prl,groupedaddress]{revtex4}

\usepackage{amsfonts}
\usepackage{graphicx}

\begin{document}


\title{Spectrum of Yang-Mills Theory in D=3+1}


\author{Marco Frasca}
\email[]{marcofrasca@mclink.it}
\affiliation{Via Erasmo Gattamelata, 3 \\ 00176 Roma (Italy)}


\date{\today}

\begin{abstract}
We give a comparison of the spectrum of Yang-Mills theory in $D=3+1$, recently derived with a
strong coupling expansion, with lattice data. We verify excellent agreement also for
2$^{++}$ glueball. A deep analogy with the $D=2+1$ case is obtained and a full quantum
theory of this approach is also given.
\end{abstract}

\pacs{11.15.Me, 11.15.-q}

\maketitle


Yang-Mills quantum theory in the low energy limit represents one of the most difficult
physical problems to face with. Difficulties in managing this theory have largely motivated
the use of large computers to get the spectrum. Recently, after a relevant theoretical
work mostly due to Nair and Karabali \cite{kn1,kn2,kn3,kn4,kn5}, it has been possible to obtain the spectrum of
this theory in 2+1 dimensions \cite{lmy}. The agreement is quite satisfactory and improvements
on this line are possible as further clarifications on the lattice data are put forward. This
approach seems promising also to manage the 3+1 case \cite{flm}.

Recently, we proposed an approach to treat the 3+1 case by a strong coupling expansion,
then we showed that this expansion is nothing else than a gradient expansion \cite{fra1,fra2,fra3,fra4,fra5,fra6}.
This approach gives the spectrum of the 0$^{++}$ glueballs in good agreement with recent
lattice computations \cite{tep1,tep2}. Furthermore we were able to give a very nice
agreement between our propagator \cite{fra3} and lattice data \cite{ste} when the gluon mass is taken to be 389 MeV 
that is the only parameter to fit. 

In this paper we have two aims. Firstly a full quantum theory is given using
functional integration. Secondly, we verify that the spectrum of the glueballs obtained
on the lattice \cite{tep1,tep2} is properly reproduced. As a by-product we obtain a deep
similarity with the aforementioned D=2+1 case.

In order to make this paper self-contained we give here a derivation of the propagator. The
starting point is a $\lambda\phi^4$ theory with the generating functional
\begin{equation}
   Z[j]=\int[d\phi]e^{\left\{i\int d^Dx\left[
   \frac{1}{2}(\partial_t\phi)^2-\frac{\lambda}{4}\phi^4+j\phi
   \right]\right\}}
   e^{\left\{-i\int d^Dx\left[\frac{1}{2}(\nabla\phi)^2+\frac{\mu_0^2}{2}\phi^2
   \right]\right\}}
\end{equation}
where we have properly separated the spatial part from the time derivatives of the field.
A gradient expansion means that we make an expansion as
\begin{equation}
    Z[j]=\exp\left[\frac{i}{2}\int d^Dy_1d^Dy_2\frac{\delta}{\delta j(y_1)}(-\nabla^2+\mu_0^2)\delta^D(y_1-y_2)
    \frac{\delta}{\delta j(y_2)}\right]Z_0[j]
\end{equation}
being
\begin{equation}
   Z_0[j]=\int[d\phi]e^{\left\{i\int d^Dx\left[
   \frac{1}{2}(\partial_t\phi)^2-\frac{\lambda}{4}\phi^4+j\phi
   \right]\right\}}.
\end{equation}
The final point is to write this latter integral in a Gaussian form. In order to realize
this point we use the fact that a small time expansion exists for the solution of the
equation
\begin{equation}
   \ddot\phi+g\phi^3=j
\end{equation}
after the change to dimensionless variables is made through $\mu_0$ as
$x\rightarrow\mu_0 x$, $\phi^2\rightarrow\mu_0^{2-D}\phi^2$ and 
introducing the coupling constant $g=\frac{\lambda}{\mu_0^{4-D}}$. The small time expansion
permits us to write the solution of this equation by a Green function given by
\begin{equation}
   \ddot G+gG^3=\delta(t)
\end{equation}
as \cite{fra7,fra8}
\begin{equation}
   \phi(t)\approx\int_0^t dt'G(t-t')j(t')+a\int_0^tdt'G(t-t')(t-t')j(t')+b\int_0^tdt'G(t-t')(t-t')^2j(t')\ldots
\end{equation}
being $a,b,\ldots$ coefficients to be computed and dependent on the initial value $j(0)$
and its derivatives. We recognize at the leading order the familiar result of Green functions
for linear systems and this implies that, as a leading approximation, one can write a
Gaussian expression for $Z_0[j]$ \cite{fra1}. In order to write this expression, let us
write down the Green function for our theory at the leading order. One has
\begin{equation}
    G(t)=\theta(t)\left(\frac{2}{g}\right)^{\frac{1}{4}}{\rm sn}\left[\left(\frac{g}{2}\right)^{\frac{1}{4}}t,i\right]
\end{equation}
and its time reversed $G(-t)$. Here one has ${\rm sn}$ a Jacobi elliptical function and
$\theta(t)$ the Heaviside function. So, one can introduce the Feynman propagator
$\Delta(x,t)=\delta^{D-1}(x)[G(t)+G(-t)]$ and write down the Gaussian expression \cite{fra1}
\begin{equation}
   Z_0[j]=\exp\left[\frac{i}{2}\int d^Dx_1d^Dx_2j(x_1)\Delta(x_1-x_2)j(x_2)\right]
\end{equation}
and then the theory can be immediately integrated giving the two-point function 
(undoing dimensionless changes)\cite{fra3}
\begin{equation}
\label{eq:prop}
    \Delta(p)=\sum_{n=0}^\infty\frac{B_n}{p^2-m_n^2-\mu_0^2+i\epsilon}
\end{equation}
being
\begin{equation}
    B_n=(2n+1)\frac{\pi^2}{K^2(i)}\frac{(-1)^{n+1}e^{-(n+\frac{1}{2})\pi}}{1+e^{-(2n+1)\pi}},
\end{equation}
and the mass spectrum of the theory given by
\begin{equation}
\label{eq:ms}
    m_n = (2n+1)\frac{\pi}{2K(i)}\left(\frac{\lambda}{2\mu_0^{4-D}}\right)^{\frac{1}{4}}\mu_0
\end{equation}
proper to a harmonic oscillator. In this case, the mass gap in the limit $\lambda\rightarrow\infty$ is
\begin{equation}
\label{eq:ds}
     \delta_S = \frac{\pi}{2K(i)}\left(\frac{\lambda}{2\mu_0^{4-D}}\right)^{\frac{1}{4}}\mu_0
\end{equation}
corresponding to the choice $n=0$, that is produced by the self-interaction of the scalar field.
In order to get eq.(\ref{eq:prop}) we have used the known relation for the Jacobi elliptical 
function \cite{gr}
\begin{equation}
    {\rm sn}(u,i)=\frac{2\pi}{K(i)}\sum_{n=0}^\infty\frac{(-1)^ne^{-(n+\frac{1}{2})\pi}}{1+e^{-(2n+1)\pi}}
    \sin\left[(2n+1)\frac{\pi u}{2K(i)}\right]
\end{equation}
being $K(i)=\int_0^{\frac{\pi}{2}}\frac{d\theta}{\sqrt{1+\sin^2\theta}}\approx 1.3111028777$ a constant.
So, the spectrum is defined through a renormalized mass $\delta_S$. The interesting case, $D=4$, gives
\begin{equation}
    m_R = \frac{\pi}{2K(i)}\left(\frac{\lambda}{2}\right)^{\frac{1}{4}}\mu_0.
\end{equation}
At this order the theory appears finite but it is possible that higher order corrections to
the Gaussian approximation could introduce the need for renormalization in this limit. 
Presently, we content ourselves with these results and we show how they apply to a
Yang-Mills theory in 3+1 dimensions.

A spectrum for the Yang-Mills theory can be obtained if we are able to map its quantum theory
on that of the scalar field discussed above. This in turn means that we have to prove that a
gradient expansion, identical to the one of the scalar field, does hold also for a Yang-Mills
theory. For our aims, we turn to the classical theory. If we are able to prove the existence
of the correspondence of the two theories in this case, the existence of the corresponding
gradient expansion for the Yang-Mills theory is obtained. This we proved in \cite{fra2}.
Indeed, the classical scalar theory has the following Hamilton equations working now in
$D=3+1$
\begin{eqnarray}
\label{eq:phi}
    \partial_t\phi &=& \pi \\ \nonumber
	\partial_t\pi  &=& \nabla^2\phi -\mu_0^2\phi -\lambda\phi^3.
\end{eqnarray}
We do the following changes and expansions
\begin{eqnarray}
   \tau &=& \sqrt{\lambda}t \\ \nonumber
   \pi &=& \sqrt{\lambda}\left(\pi^{(0)} + \frac{1}{\lambda}\pi^{(1)}+ \frac{1}{\lambda^2}\pi^{(2)} + \ldots\right) \\ \nonumber
   \phi &=& \phi^{(0)} + \frac{1}{\lambda}\phi^{(1)} + \frac{1}{\lambda^2}\phi^{(2)} + \ldots.
\end{eqnarray}
and then we get the following non-trivial equations for our expansions
\begin{eqnarray}
\label{eq:phis}
    \partial_{\tau}\phi^{(0)} &=& \pi^{(0)} \\ \nonumber
	\partial_{\tau}\phi^{(1)} &=& \pi^{(1)} \\ \nonumber
    \partial_{\tau}\phi^{(2)} &=& \pi^{(2)} \\ \nonumber
	                 &\vdots&  \\ \nonumber
	\partial_{\tau}\pi^{(0)} &=& -\phi^{(0)3} \\ \nonumber
	\partial_{\tau}\pi^{(1)} &=& -3\phi^{(0)2}\phi^{(1)} + \nabla^2\phi^{(0)}-\mu_0^2\phi^{(0)} \\ \nonumber
	\partial_{\tau}\pi^{(2)} &=& -3\phi^{(0)}\phi^{(1)2}-3\phi^{(0)2}\phi^{(2)} + \nabla^2\phi^{(1)}-\mu_0^2\phi^{(1)}\\ \nonumber
	                 &\vdots& \label{eq:pert}
\end{eqnarray}
that we recognize as a gradient expansion as should be. These equations can be solved and
give a meaningful gradient expansion for a classical $\lambda\phi^4$ theory. We can proceed
in a similar way for a SU(N) Yang-Mills field. The Hamilton equations are,
in the gauge $A_0^a=0$ that needs to be fixed to carry on our computation, \cite{fs,smi}
\begin{eqnarray}
    \partial_t A_k^a&=&F_{0k}^a \\ \nonumber
    \partial_t F_{0k}^a&=&\partial_lF_{lk}^a+gf^{abc}A_l^bF_{lk}^c
\end{eqnarray}
being $g$ the coupling constant, $f^{abc}$ the structure constants of the gauge group,
$F_{lk}^a=\partial_lA_k^a-\partial_kA_l^a+gf^{abc}A_l^bA_k^c$ and the constraint
$\partial_kF_{0k}^a+gf^{abc}A_k^bF_{0k}^c=0$ does hold. So, let us introduce the
following equations, as done for the scalar field,
\begin{eqnarray}
   \tau &=& gt \\ \nonumber
   F_{0k}^a&=& gF_{0k}^{a(0)} + F_{0k}^{a(1)} + \frac{1}{g}F_{0k}^{a(2)} + \ldots \\ \nonumber
   F_{lk}^a&=& F_{lk}^{a(0)} + \frac{1}{g}F_{lk}^{a(1)} + \frac{1}{g^2}F_{lk}^{a(2)} + \ldots \\ \nonumber
   A_k^a &=& A_k^{a(0)} + \frac{1}{g}A_k^{a(1)} + \frac{1}{g^2}A_k^{a(2)} + \ldots.
\end{eqnarray}
Then, one has the perturbation equations
\begin{eqnarray}
    \partial_\tau A_k^{a(0)}&=&F_{0k}^{a(0)} \\ \nonumber
    \partial_\tau A_k^{a(1)}&=&F_{0k}^{a(1)} \\ \nonumber
    &\vdots& \\ \nonumber
    \partial_\tau F_{0k}^{a(0)}&=&f^{abc}f^{cde}A_l^{b(0)}A_l^{d(0)}A_k^{e(0)} \\ \nonumber
    \partial_\tau F_{0k}^{a(1)}&=&
    f^{abc}f^{cde}A_l^{b(1)}A_l^{d(0)}A_k^{e(0)}
    +f^{abc}f^{cde}A_l^{b(0)}A_l^{d(1)}A_k^{e(0)}
    +f^{abc}f^{cde}A_l^{b(0)}A_l^{d(0)}A_k^{e(1)} \\ \nonumber
    & &+f^{abc}\partial_l\left(A_l^{b(0)}A_k^{c(0)}\right)
    +f^{abc}A_l^{b(0)}\left(\partial_lA_k^{c(0)}-\partial_kA_l^{c(0)}\right) \\ \nonumber
    &\vdots&
\end{eqnarray}
and we can recognize at the leading order the homogeneous Yang-Mills equations. These equations
display a rich dynamics as e.g. Hamiltonian chaos \cite{sav1,sav2,sav3}. But chaotic solutions
do not give a meaningful quantum field theory. But we easily realize that taking all the
components of the Yang-Mills field as equal \cite{smi} we can make a complete correspondence between
the gradient expansion of the classical scalar field (\ref{eq:phis}) putting $\lambda=Ng^2$,
the so called t'Hooft coupling. An arbitrary mass is needed in this case because no mass
scale exists in the Yang-Mills theory. This is an integration constant in this theory
and is generally obtained, also on the lattice by fixing the string strength, by
comparison with experimental data. In the following we will fix this constant to \cite{tep1,tep2}
\begin{equation}
     \sqrt{\sigma}=\left(\frac{Ng^2}{2}\right)^{\frac{1}{4}}\Lambda={\rm 440\ MeV}.
\end{equation}
This choice, as we will se below, permits us to reproduce exactly the lattice spectrum.  
So, as this series exists for the scalar field it does
exist for the Yang-Mills field \cite{fra2}. We used the fact the $g$ in $D=3+1$ is
dimensionless as happens to $\lambda$ for the scalar field and we have used it as a
useful placeholder for our expansion. But the same result could be obtained by removing
it from the theory using the duality principle in perturbation theory \cite{fra9}.

Now, in order to complete our proof, we consider this procedure starting from the
functional integral of the Yang-Mills theory. One has, choosing now the gauge $\partial_\mu A^{a\mu}=0$, \cite{nair}
\begin{equation}
    Z_{YM}[j]=\int [dA][dc][d\bar c]e^{iS_0+iS_{INT}}
\end{equation}
being
\begin{equation}
    S_0=\int d^4x\left[\frac{1}{2}\partial_\mu A^a_\nu\partial^\mu A^{a\nu}+\partial^\mu\bar c^a\partial_\mu c^a\right]
\end{equation}
and
\begin{equation}
    S_{INT}=\int d^4x\left[gf^{abc}\partial_\mu A_\nu^aA^{b\mu}A^{c\nu}
	+\frac{g^2}{4}f^{abc}f^{ars}A^b_\mu A^c_\nu A^{r\mu}A^{s\nu}
	+gf^{abc}\partial_\mu\bar c^a A^{b\mu}c^c+j^{a\mu}A^a_\mu\right].
\end{equation}
We can assume for $A_\mu^a$ that one of the spatial components is zero and the others are all
equal in agreement with the classical case. This is not a truncated Yang-Mills theory but a
proof of the existence of a quantum solution completely mapping a $\lambda\phi^4$ theory that
holds in the infrared. For consistency reasons we do the same for the ghost field that in this
way happens to decouple from the Yang-Mills field and obtaining in this way a propagator
going like $1/p^2$ that is diverging in agreement with lattice computations\cite{ste}. So, finally
one has
\begin{equation}
    S_0=-(N^2-1)\int d^4x\left[\frac{1}{2}\partial_\mu A\partial^\mu A +\partial\bar c\partial c\right]
\end{equation}
and
\begin{equation}
    S_{INT}=-(N^2-1)\int d^4x\left[\frac{Ng^2}{4}A^4+jA\right]
\end{equation}
where we note the factor $N^2-1$ needed to preserve the number of degrees of freedom. Then,
a gradient expansion holds also for the quantum case and the given theory exists being the same as
for a $\lambda\phi^4$ quantum field theory in the infrared. Having a zero mass theory,
scale invariance is retained as for the Yang-Mills theory. Then,  
as for the classical case, we need an arbitrary mass scale to normalize the theory. In
this case, as noted above, we call this mass $\Lambda$.
 
The above correspondence in a gradient expansion between a scalar field theory in $D=3+1$ and
a Yang-Mills theory gives us all the information we need about the spectrum and the
propagator of the latter. Indeed we have
\begin{equation}
\label{eq:prop1}
    \Delta_{YM}(p)=\sum_{n=0}^\infty\frac{B_n}{p^2-M(n)^2+i\epsilon}
\end{equation}
being
\begin{equation}
    B_n=(2n+1)\frac{\pi^2}{K^2(i)}\frac{(-1)^{n+1}e^{-(n+\frac{1}{2})\pi}}{1+e^{-(2n+1)\pi}},
\end{equation}
and the mass spectrum of the theory given by
\begin{equation}
\label{eq:ms1}
     \frac{M(n)}{\sqrt{\sigma}}= (2n+1)\frac{\pi}{2K(i)}.
\end{equation}
The mass gap is
\begin{equation}
\label{eq:ds1}
     M(0) = \frac{\pi}{2K(i)}\sqrt{\sigma}.
\end{equation}
to be identified with the gluon mass dynamically acquired by the strong interaction 
due to the self-interaction terms of the Yang-Mills field.
We note that, differently from the case $D=2+1$, the string strength is an arbitrary
constant that should be derived by experiment or a higher order theory. Indeed, we are not
able to apply this correspondence to a lower dimensional Yang-Mills theory, this is indeed
a lucky case, and then the approach given in \cite{kn1,kn2,kn3,kn4,kn5,lmy,flm} is welcome
as could be its application to a $D=3+1$ case to be compared with our results.

We now compare our results with lattice data \cite{tep1,tep2,ste}. For the propagator
we just note that the agreement is excellent fixing the gluon mass to 389 MeV \cite{fra3}.
We do not pursuit this matter further here waiting for other data from lattice computations.
We note that we have obtained a result that does not depend on $N$, the group order. This
is aligned with lattice results as given in \cite{tep1,tep2}. So, for our convenience
we limit the comparison to the continuum for the SU(3) group. Lattice data in \cite{tep1,tep2}
are given only for 0$^{++}$, 0$^{++*}$ and 2$^{++}$ and we assume these results are the ones
to be more confident. Further results for the spectrum can be read in \cite{mor} and
\cite{mey}, the latter being an unofficial extension of the results given in \cite{tep1,tep2}.
We do not aim to go that far but just to have a proper understanding of how to get, given
the propagator, higher order excitations. Initially we just notice that $M(n)$ is indeed
the spectrum of 0$^{++}$. The reason for this relies on the fact that these are poles of
the propagator and represent physical states of the theory. So, we have to interpret them
in this way and indeed we obtain agreement with lattice computations. 
In Tab. \ref{tab:0++} we can see an exceedingly good agreement
between lattice and theoretical data.
\begin{table}
\begin{tabular}{|c|c|c|c|} \hline\hline
Excitation & Lattice & Theoretical & Error \\ \hline
$\sigma$   & -       & 1.198140235 & - \\ \hline 
0$^{++}$   & 3.55(7) & 3.594420705 & 1\% \\ \hline
0$^{++*}$  & 5.69(10)& 5.990701175 & 5\% \\ \hline
0$^{++**}$ & -      & 8.386981645 & - \\ \hline\hline
\end{tabular}
\caption{\label{tab:0++} Comparison for the 0$^{++}$ glueball}
\end{table}
The interesting point is about the excitation called $\sigma$ in our table. We called
this particle in this way as $M(0)=527\ MeV$ that is about the mass of the $\sigma$
resonance in \cite{pdg}. This is another theoretical result of our approach and currently
it is not expected to be seen in lattice computations. We expect to see its excited
states in higher excited states of the spectrum.

Our aim and the main motivation of this paper is to see what happens when higher excited
states than 0$^{++}$ are considered. This analysis can be carried out as already done in
D=2+1 \cite{lmy}. We consider a time independent correlator given from the propagator (\ref{eq:prop1})
by taking $p_0=0$ and Fourier transforming in space obtaining
\begin{equation}
   C(x,y)=\sum_{n=0}^{+\infty}\frac{A_n}{|x-y|}e^{-M(n)|x-y|}
\end{equation}
that has an identical form to the asymptotic correlator obtained in D=2+1 for 0$^{++}$ states
as should be\cite{lmy}. Taking the square of this expression we obtain for the 2$^{++}$ spectrum
\begin{equation}
   \frac{M_2(n,m)}{\sqrt{\sigma}}=\frac{M(n)+M(m)}{\sqrt{\sigma}}=(2n+2m+2)\frac{\pi}{2K(i)}
\end{equation}
that gives Tab.\ref{tab:2++} through $M_2(0,1)=M_2(1,0)$. 
We observe a degeneracy, the same of the two-dimensional harmonic oscillator.
\begin{table}
\begin{tabular}{|c|c|c|c|} \hline\hline
Excitation & Lattice & Theoretical & Error \\ \hline
$\sigma^*$ & -       & 2.396280470 & - \\ \hline 
2$^{++}$   & 4.78(9) & 4.792560940 & 0.2\% \\ \hline
2$^{++*}$  & -       & 7.188841410 & - \\ \hline\hline
\end{tabular}
\caption{\label{tab:2++} Comparison for the 2$^{++}$ glueball}
\end{table}
The agreement is again astonishingly good.

A few words should be spent about the strong similarities between the D=2+1 and D=3+1 cases.
In both cases we observe the same asymptotic form of the propagator and the same harmonic
oscillator spectrum that gives exceedingly good agreement with lattice data. As a final note
we point out that some analysis on the lattice support the view that the spectrum of
the theory in $D=2+1$ is indeed that of a harmonic oscillator \cite{mck1,mck2}.

We have given a full account of a strong coupling approach that gives the spectrum of
a Yang-Mills theory in D=3+1. The agreement with lattice data are exceedingly good and
the analogy with the Yang-Mills theory in D=2+1 is really striking putting our results
on a sound ground. Finally we have also given a fully quantum field formulation of
Yang-Mills theory in our approach.



\begin{thebibliography}{99}
\bibitem{kn1} D. Karabali and V. P. Nair, Nucl. Phys. B {\bf 464}, 135 (1996). 
\bibitem{kn2} D. Karabali and V. P. Nair, Phys. Lett. B {\bf 379}, 141 (1996). 
\bibitem{kn3} D. Karabali, C. J. Kim and V. P. Nair, Nucl. Phys. B {\bf 524}, 661 (1998). 
\bibitem{kn4} D. Karabali, C. J. Kim and V. P. Nair, Phys. Lett. B {\bf 434}, 103 (1998). 
\bibitem{kn5} D. Karabali, C. J. Kim and V. P. Nair, Phys. Rev. D {\bf 64}, 025011 (2001).
\bibitem{lmy} R. G. Leigh, D. Minic and A. Yelnikov, hep-th/0604060.
\bibitem{flm} L. Freidel, R. G. Leigh and D. Minic, Phys. Lett. B {\bf 641}, 105 (2006).
\bibitem{fra1} M. Frasca, Phys. Rev. D {\bf 73}, 027701 (2006); Erratum-ibid., 049902 (2006).
\bibitem{fra2} M. Frasca, Int. J. Mod. Phys. A {\bf 22}, 1727 (2007).
\bibitem{fra3} M. Frasca, Int. J. Mod. Phys. A {\bf 22}, 2433 (2007).
\bibitem{fra4} M. Frasca, Int. J. Mod. Phys. A {\bf 22}, 1441 (2007).
\bibitem{fra5} M. Frasca, hep-th/0511173.
\bibitem{fra6} M. Frasca, Int. J. Mod. Phys. D {\bf 15}, 1373 (2006).
\bibitem{tep1} B. Lucini, M. Teper, Phys. Rev. D {\bf 64}, 105019 (2001).
\bibitem{tep2} B. Lucini, M. Teper, U. Wenger, JHEP {\bf 06}, 012 (2004).
\bibitem{ste} A. Sternbeck, E.-M.Ilgenfritz, M. M\"uller-Preussker, A. Schiller, I. L. Bogolubsky, PoS(LAT2006)076.
\bibitem{fra7} M. Frasca, arXiv:0704.1568 [hep-th].
\bibitem{fra8} M. Frasca, Mod. Phys. Lett. A {\bf 22}, 1293 (2007).
\bibitem{gr} I. S. Gradshteyn, I. M. Ryzhik, {\sl Table of Integrals, Series, and Products}, (Academic Press, 2000).
\bibitem{fs} L. D. Fadeev and A. A. Slavnov, {\sl Gauge Fields, Introduction to Quantum Theory}
(Benjamin-Cummings, Reading, 1980).
\bibitem{smi} A: V. Smilga, {\sl Lectures in Quantum Chromodynamics}, (World Scientific, Singapore, 2001).
\bibitem{sav1} S. G. Matinyan, G. K. Savvidy, N. G. Ter-Arutunian Savvidy, Sov. Phys. JETP {\bf 53}, 421 (1981).
\bibitem{sav2} G. K. Savvidy, Phys. Lett. B {\bf 130}, 303 (1983).
\bibitem{sav3} G. K. Savvidy, Nucl. Phys. B {\bf 246}, 302 (1984).
\bibitem{fra9} M. Frasca, Phys. Rev. A {\bf 58}, 3439 (1998).
\bibitem{nair} V. P. Nair, {\sl Quantum Field Theory}, (Springer, New York, 2005).
\bibitem{mor} Y. Chen, A. Alexandru, S. J. Dong, T. Draper, I. Horvath, F. X. Lee, K. F. Liu, N. Mathur, 
C. Morningstar, M. Peardon, S. Tamhankar, B. L. Young, J. B. Zhang, Phys. Rev. D {\bf 73}, 014516 (2006).
\bibitem{mey} H. Meyer, hep-lat/0508002, Ph.D. Thesis.
\bibitem{pdg} W.-M. Yao et al., J. Phys. G {\bf 33}, 1 (2006).
\bibitem{mck1} B. H. J. McKellar, J. Carlsson, Nucl.\ Phys.\ Proc.\ Suppl.\  {\bf 129}-{\bf 130}, 420 (2004).
\bibitem{mck2} B. H. J. McKellar, J. Carlsson, Nucl.\ Phys.\ Proc.\ Suppl.\  {\bf 141}, 179 (2005). 
\end{thebibliography}
\end{document}